\documentclass[pre,twocolumn,showpacs,preprintnumbers,floatfix,english]{revtex4}

\usepackage[utf8]{inputenc}
\usepackage[T1]{fontenc}
\usepackage{babel}

\usepackage{ifpdf}
\usepackage{graphicx}
\usepackage{color}
\usepackage{lineno}
\usepackage{pslatex}

\usepackage{bm}
\usepackage{dcolumn}

\usepackage{amsmath}
\usepackage{amssymb}
\usepackage{amscd}

\ifpdf
\usepackage{epstopdf}
\usepackage[%
  pdftitle={lGe2Se3},%
  pdfauthor={},%
  pdfsubject={lGe2Se3},%
  pdfkeywords={lGe2Se3},%
  pdfstartview=FitH,%
  bookmarks=true,%
  bookmarksopen=true,%
  breaklinks=true,%
  colorlinks=true,%
  linkcolor=blue,anchorcolor=blue,%
  citecolor=blue,filecolor=blue,%
  menucolor=blue,pagecolor=blue,%
  urlcolor=blue]{hyperref}
\else
\usepackage[%
  pdftitle={lGe2Se3},%
  pdfauthor={},%
  pdfsubject={lGe2Se3},%
  pdfkeywords={lGe3Se3},%
  breaklinks=true,%
  colorlinks=true,%
  linkcolor=blue,anchorcolor=blue,%
  citecolor=blue,filecolor=blue,%
  menucolor=blue,pagecolor=blue,%
  urlcolor=blue]{hyperref}
\fi

\include{bibsymb}

\newcommand{\vo}{V$_\text{2}$O$_\text{5}$}

\begin{document}
\draft

\title{Designing heavy metal oxide glasses with threshold properties from network rigidity}

\author{Shibalik Chakraborty$^1$, P. Boolchand$^1$, M. Malki$^{2,3}$, M. Micoulaut$^4$}

\affiliation{$^1$School of Electronics and Computing Systems, College of Engineering and Applied Science, University of Cincinnati
Cincinnati, OH 45221-0030, USA}

\affiliation{$^2$ CEMHTI, CNRS UPR 3079, 1D, Avenue de la Recherche Scientifique, 45071 Orle\'ans Cedex 02, France}
\affiliation{$^3$ Universit\'e d’Orléans (Polytech’ Orl\'eans), BP 6749, 45072 Orl\'eans Cedex 02, France}
\affiliation{$^4$Laboratoire de Physique Th\'{e}orique de la Mati\`ere Condensée,
Universit\'{e} Pierre et Marie Curie, 4 Place Jussieu, F-75252 Paris Cedex 05, France}

\date{\today}

\begin{abstract}
Here we show that a new class of glasses composed of heavy metal oxides involving transition metals (V$_\text {2}$O$_\text{}$5-TeO$_\text{2}$), can surprisingly be designed from very basic tools using topology and rigidity of their underlying molecular networks. When investigated as a function of composition, such glasses display abrupt changes in network packing and enthalpy of relaxation at Tg, underscoring presence of flexible to rigid elastic phase transitions. We find that these elastic phases are fully consistent with polaronic nature of electronic conductivity at high V$_\text{2}$O$_\text{5}$ content. Such observations have new implications for designing electronic glasses which differ from the traditional amorphous electrolytes having only mobile ions as charge carriers. 
\end{abstract}


\maketitle

Only one element of the Periodic Table (Se) vitrifies at ambient pressure but a large number of glassy alloys can be obtained by combining appropriate elements from Group III, IV and V with Group VI elements (O, S, Se and Te) such as silica (SiO$_\text{2}$) or germanium diselenide (GeSe$_\text{2}$). In the design of physical, chemical or electrical properties, these base network glasses are then alloyed with appropriate modifiers (e.g. Na$_\text{2}$O) or additives (AgI) \cite{r1}. Molecular structure of the base network glasses is a prerequisite to develop a basic understanding of a structure-property relationship. The optimization of functionalities upon modification with alkali–oxides or electrolyte additives is then realized on an empirical basis, given the fact that the number of possible combinations is infinite. Another challenge is the
absence of any precise information on the microscopic origin of the glass-forming ability itself, the resistance towards crystallization and the kinetics of glass transition. Therefore, few attempts have really been made to obtain glasses from important additives such as transition elements. As a result, the latter usually appear as minor additives for selected applications. For instance, Titania (TiO$_\text{2}$) with less than one mole percent is used for photo-active applications \cite{r2} such as self-cleaning coatings while Chromia (Cr$_\text{2}$O$_\text{3}$) has been used for centuries as a pigment for the green coloring of silicate glasses \cite{r3}. 
\par
Can network glasses be realized with large fraction of transition metal-oxides ? This knowledge appears to be essential for new applications that could emerge from detailed compositional studies, which would open an avenue to ultimately tune materials synthesis to specific applications. Here, we study thermal properties of heavy metal oxide glasses (HMO), composed of TeO$_\text{2}$ and \vo.Telluria and Vanadia as such are bad glass formers, but, their pseudo-binary alloys form excellent bulk glasses over wide proportions (18\%$<$x$<$55\%). 
\par
What is the origin of this remarkable behavior?  
\par
Currently this is not known. Here we show for the first time that the glass forming ability in this family of HMO is topological in origin, and fundamentally derives from the onset of network rigidity. We arrive at the result by establishing a Maxwell stability criterion for isostatic rigid networks, since the nature of local structures of both TeO$_\text{2}$ and V$_\text{2}$O$_\text{5}$, and their evolution with composition are available from detailed NMR experiments \cite{r4}. Results show that TeO$_\text{2}$-V$_\text{2}$O$_\text{5}$ networks become optimally constrained near 20 mole \% of V$_\text{2}$O$_\text{5}$, and display experimentally all the features characteristic of two elastic phase transitions noted earlier in several network glasses; a stress-transition (stressed-rigid to rigid near x = 23.5\%) followed by a rigidity-transition (rigid to flexible near x = 26.0\%). The observation, to the best of our knowledge, is the first of its kind in a HMO. In addition, we find that electronic conductivity displays three regimes of variation with a threshold in electronic conduction near the rigidity transition (x = 26.0\%). Since electron transport is strongly tied to a local network deformation, one identifies its origin to be polaronic in character; it is promoted by a softening of the alloyed glass network once it enters the flexible phase with high vanadia content. It is believed that these results must be generic to the family of HMO containing the transition metal oxide V$_2$O$_5$. These new results open for the first time a promising extension of concepts of topology and rigidity from chalcogenides and modified oxides to the case of HMO. In this context, the approach of topology and rigidity to decoding the functionality of chalcogenides-, and oxide- glasses has already proved to be remarkably successful \cite{r5,r6}.
\par
\begin{figure*}
\includegraphics*[width=0.8\linewidth]{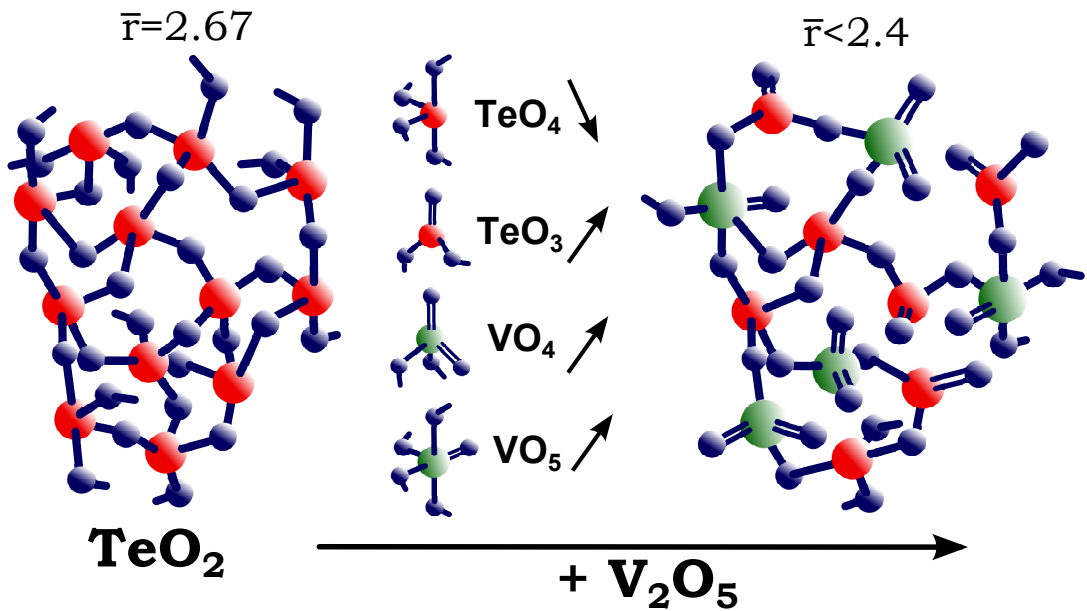}
\caption{Structural changes in heavy metal glasses. A schematic view of the structural changes induced by the Vanadia into a base Telluria glass network, following Ref. \onlinecite{r4}. The basic TeO$_\text{4}$ trigonal bipyramid (number of topological constraints n$_c$=3.67, or mean coordination number, $\bar r$=2.67) converts partially into O=TeO$_\text{2/2}$ trigonal pyramids TeO$_\text{3}$ (n$_c$=1.85, $\bar r$=2.34) while Vanadia is present either in a tetrahedral VO$_\text{4}$ geometry (n$_c$=2.5, $\bar r$=2.5) or as a 0=VO$_\text{4/2}$(VO$_\text{5}$) trigonal bipyramid (n$_c$=3.5, $\bar r$=3). The average coordination number of the base network ($\bar r$=2.67) is obviously reduced as the vanadia content increases.
}
\label{snap}
\end{figure*}
\begin{figure}
\includegraphics*[width=0.9\linewidth]{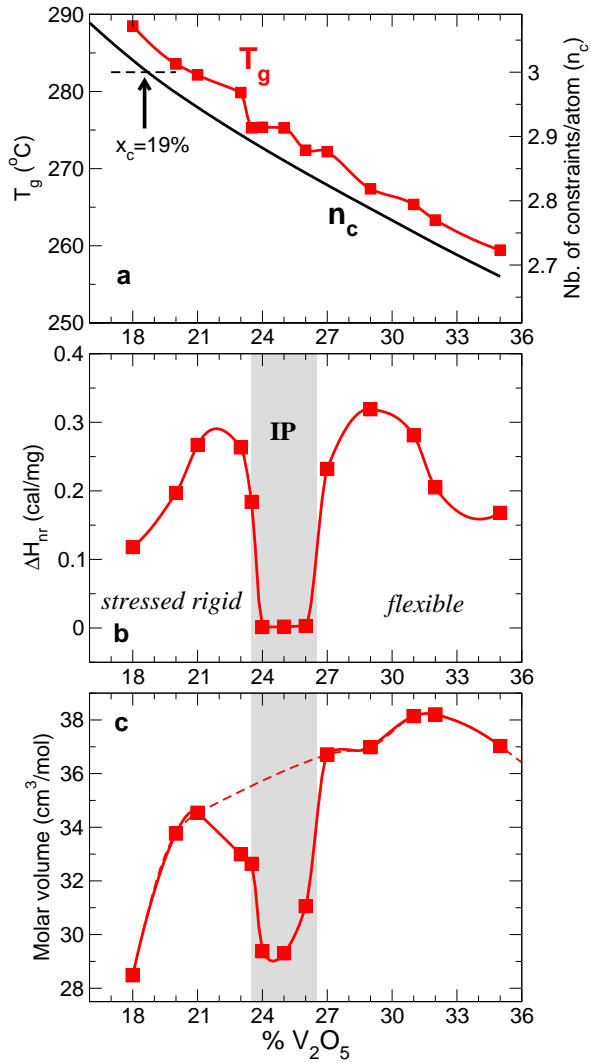}
\caption{Thermal properties of heavy metal glasses. Glass transition temperature (a), number of topological constraints (a, right axis), non-reversing heat flow $\Delta$ H$_{nr}$ (b) and molar volume variation (c) in x\vo-(95-x)TeO$_\text{2}$-5 B$\text{2}$O$_\text{3}$ glasses as a function of Vanadia content x. The square well in thermal properties defines a reversibility window, similar to the ones in chalcogenide glasses. The glass transition variation (a) can be fitted (red curve) from the number of floppy modes f (right axis, n$_c$=3-f) in the network using a Lindemann criterion in combination with rigidity theory \cite{r7}. The arrow in a) marks the Maxwell isostatic criterion determined from the species population of Sakida et al. \cite{r4}.
}
\label{calor}
\end{figure}

To support these claims, we first consider the network structure of x\vo -(95-x)TeO$_\text{2}$-5B$_\text{2}$O$_\text{3}$, the presence of a few (5) mole \% of B$_\text{2}$O$_\text{3}$ into the Telluro-Vanadate binary glasses  suppresses any tendency of segregation (see Supplementary Material). An investigation from NMR \cite{r4} has shown that the increase of \vo\ content leads to a systematic changes in network connectedness (Fig. \ref{snap}), which results from a change in the nature of building blocks. Specifically, the basic local structure of telluria composed of oxygen bridging TeO$_\text{4/2}$ trigonal bipyramids, is in fact progressively converted into a O=TeO$_\text{2/2}$ one, a quasi-trigonal Te with two bridging and one terminal (double bonded) oxygen, whereas the additive vanadia leads vanadium to exist in two environments: a O=VO$_\text{4/2}$ trigonal bipyramid with five oxygen-neighbors, and a VO$_\text{4}$ quasi-tetrahedra having two double bonded V=0 bonds (Fig. \ref{snap}). The concentrations of these building blocks depend on the \vo\ composition and converge at high vanadia content (x> 30\%) to the limit of 50\% each \cite{r4}. These local structure results lead to the network connectivity to be systematically altered by the increased presence of terminal (double-bonded) oxygen; network mean coordination number ($\bar r$) decreases from $\bar r$ = 2.67 at x=0 to $\bar r$ = 2.45 at x=80\% \vo, a feature that is directly reflected in the compositional trend of the glass transition temperature T$_g$(x) which usually reflects the change in network connectivity\cite{EPL}. Calorimetric measurements show T$_g$(x) to monotonically decrease from 290$^o$C at x = 18\% to 260$^o$C at x = 35\% (Figure \ref{calor}a). These findings provide the first clue that networks describing such glasses become steadily less connected as the Vanadia content ‘x’ increases. 
\par
Using the species population determined by Sakida et al. \cite{r4}, we then proceed to enumerate the count of mechanical constraints n$_c$(x) in these Telluro-Vanadate networks. For an atom possessing a coordination number r, the count of bond-stretching constraints (r/2), and of bond-bending constraints (2r-3), for each atom possessing a coordination number of r$\geq$ 2 allows obtaining the total n$_c$(x).   For terminal atoms, such as a double bonded terminal Oxygen in a TeO$_\text{3}$ unit, r = 1, only a bond-stretching constraint of 1/2 per atom is counted, according to rules established earlier \cite{r6b,r6t}. We have thus obtained the global count of constraints n$_c$(x) as a function of Vanadia in the Telluro-Vanadate networks, and these results are plotted in Fig. \ref{calor}a, right axis. It is found that n$_c$(x) starting from a value of about 3.50 near x = 5\% steadily decreases to 2.70 near x = 35\%.  These constraint estimates show that formation of the less connected, and therefore less constrained TeO$_\text{3}$ and VO$_\text{4}$ units at the expense of TeO$_\text{4}$ and VO$_\text{5}$ units lowers the global connectivity of these networks. Remarkably, the condition, n$_c$ = 3, for rigidity percolation is realized when x = 19\% (arrow in Fig. \ref{calor}a), the Maxwell stability criterion for isostatic structures. These findings strongly suggest that the origin of glass formation in the present HMO is driven by the optimally coordinated nature of these networks. We, thus, view networks at x $<$ 19\% to be in the stressed-rigid phase (n$_c>$ 3), while those at x > 19\% to be in the flexible (n$_c<$ 3) phase, and the composition of x = x$_c$ = 19\% to be near the phase boundary between flexible and stressed-rigid glasses.
\par
Futhermore, calorimetric measurements also reveal that the non-reversing enthalpy of relaxation ($\Delta$H$_{nr}$(x)) at T$_g$ varies in a highly non-monotonic fashion (Figure \ref{calor}b) with glass composition x; the term sharply drops and becomes minuscule as x $>$ x$_c$(1) = 23.5\%, and remains so until x $>$ x$_c$(2) = 26.5\% , when it suddenly increases  back up to a value observed below the first threshold (x$_c$(1)). The square-well like behavior of the non-reversing heat flow constitutes signature of the glass transition becoming thermally reversing in the x$_c$(1)$<$ x $<$ x$_c$(2) composition range, similar to previous findings \cite{r8,r9} reported in more common glasses. Based on these previous examples, it is a clear that the present HMO glasses are stressed-rigid at high Telluria content, but become flexible at high Vanadia content, consistent with a reduction of network connectivity.
Compositional trends in molar volumes of glasses (Figure \ref{calor}c) also show a non-monotonic variation, and decrease by about 17\% upon a few percent added \vo. Molar volumes, in general, increase with increasing Vanadia content x of glasses,  but are found to reduce in a spectacular fashion as x > 23\% and to remain low up till x $>$ 26\% when they increase sharply to acquire a value that appeared to be a smooth extrapolation (broken line in Fig. \ref{calor}b) from the low values of x $<$ 23\%. The almost square-well like drop of about 12\% in V$_m$(x) in this compositional range, appears to nearly coincide with the reversibility window obtained from calorimetric measurements. A parallel behavior has been noted earlier \cite{r5} in chalcogenides (Ge-Se) glasses, and is supportive of the space filling nature of networks \cite{Bourgel} formed in that special isostatic (n$_c\simeq$ 3) compositional window where stress is absent. 
\par
\begin{figure}
\includegraphics*[width=0.9\linewidth]{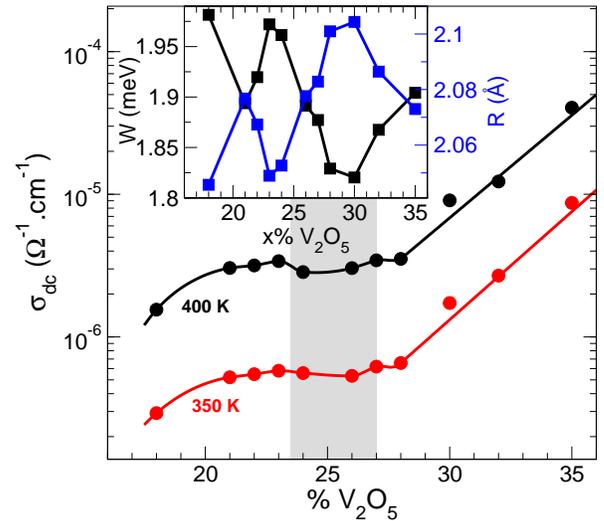}
\caption{Conductivity of heavy metal glasses. DC conductivity measurements (at 350 K and 400 K) in x\vo-(95-x)TeO$_\text{2}$-5 B$\text{2}$O$_\text{3}$ glasses as a function of Vanadia (V2O5) content x. The insert shows activation energies W(x) for polaron hopping energy, and corresponding hopping range R(x) using Mott’s variable range hopping (VRH) equation  (blue symbols, right axis).
}
\label{condux}
\end{figure} 

Finally, we provide measurements of the dc conductivity ($\sigma_{dc}$) (Figure \ref{condux}) at different temperatures. All display the same trend with Vanadia composition, i.e. a mild increase at x$<$22-23\%, a plateau behavior in the reversibility window, and a power-law increase of $\sigma\simeq$ (x-x$_c$(2))$^p$ once x$>$x$_c$(2)=27\%. Clearly, a substantial increase of electrical conduction is linked to the flexible nature of the network; we observe an increase by an order of magnitude with the addition of just 5-10\% \vo\ in the flexible elastic phase (x $>$ 27\%). As there are obviously no ions involved, sources contributing to the dc conduction mechanism can either arise from electrons or from polarons \cite{r10}. However, the measurement of the low temperature (300 K) relaxation of the measured dielectric constant ($\epsilon^{'}\simeq$40-60) appears to be activated with a frequency of 0.1 MHz, a frequency which is by far too low to be associated with an electronic tunneling mechanism \cite{r11}. The only possible mechanism for charge transport is therefore polaronic in nature, which is, indeed, consistent with the presently established flexible nature of the network at high vanadia content, as local network polarization and deformation or distortion caused by a moving electron is facilitated by the presence of floppy modes \cite{r7} in the elastically flexible (at x$>$27\%) phase. The observation is also consistent with the fact that Mott’s variable-range-hopping (VRH) equation for polaronic conductivity \cite{r12,r13} of the form $\sigma_{dc}=\sigma_0$exp[-(T$_0$/T)$^{1/4}$] fits better the low temperature $\sigma_{dc}$ experimental data than a standard Arrhenius equation $\sigma_{dc}=\sigma_0$exp[-E$_A$/k$_B$T] (see Supplementary Material), thus strongly confirming the polaronic nature of conduction in the flexible glasses.
Furthermore, T$_0$ in the Mott VRH formalism is related to the hopping energy W and hopping range R of polarons via:
\begin{eqnarray}
T_0={\frac {24}{\pi k_BN(E_F)D^3}}
\end{eqnarray}
\begin{eqnarray}
W={\frac {3}{4\pi R^3N(E_F)}}
\end{eqnarray}
\begin{eqnarray}
R^2=\sqrt{\frac {D}{8\pi N(E_F)k_BT}}
\end{eqnarray}
where $D$ is the decay length of the localized wave-functions, usually the distance between two neighboring atoms, and $N$($E_F$) is the density of localized states at the Fermi level. The distance D is here taken equal to 2.76~\AA\ which is the distance between two neighboring oxygen atoms, the typical distanceinvolved in Vanadates \cite{r18}. Both quantities $W$ and $R$ appear to capture as well (inset of Fig. \ref{condux}) the softening of the glass network by displaying a steep decrease of W(x) and a steep increase of R(x) in the reversibility window.
\par
Global consequences of the findings can now be enunciated. In a fashion very similar to those used for common modified oxide and chalcogenide network glasses, it is the first time that ideas on topology and network connectivity have been used here to  understanding of glass-forming tendency in HMO, a family of glasses that have received much less attention up to now. A flexible to rigid transition is found in these telluro-vanadates which bears striking similarities with those found in other common glasses including commercial ones. For the first time, we show that an electronic transition takes place, which manifests in a power-law increase of the dc conductivity as the Vanadia content exceeds a certain threshold value (27\%). This transition is deeply connected to the rigid to flexible transition induced by the loss of connectivity as local structures with lower connectedness appear in the networks. Electrical conductivity is polaronic in character, and increases precipitously due to floppy modes that facilitate electron transport through local deformations in the elastically flexible phase.

\end{document}